\documentclass[journal=jpclcd,manuscript=letter]{achemso}

\usepackage[version=3]{mhchem} %
\usepackage[T1]{fontenc}       %
\usepackage{amsmath}
\usepackage{graphicx}
\usepackage{caption}
\usepackage{subcaption}
\usepackage{txfonts}
\usepackage{xr}
\usepackage[english]{babel}
\usepackage[usenames]{color}
\usepackage[sort&compress,numbers,super]{natbib}
\setcitestyle{square}

\makeatletter
\newcommand*{\addFileDependency}[1]{%
  \typeout{(#1)}
  \@addtofilelist{#1}
  \IfFileExists{#1}{}{\typeout{No file #1.}}
}
\makeatother
 
\newcommand*{\myexternaldocument}[1]{%
    \externaldocument{#1}%
    \addFileDependency{#1.tex}%
    \addFileDependency{#1.aux}%
}
\usepackage[shortcuts]{extdash}

\myexternaldocument{SupportingInformation}

\author{Riccardo Dettori}
\email{rdettori@ucdavis.edu}
\affiliation[UCDavis]
{Department of Chemistry, University of California Davis, One Shields Avenue, Davis, California 95616, United States}
\author{Michele Ceriotti}
\affiliation[École Polytechnique Fédérale de Lausanne]
{Laboratory of Computational Science and Modeling, IMX, \'Ecole Polytechnique F\'ed\'erale de Lausanne, 1015 Lausanne, Switzerland}
\author{Johannes Hunger}
\affiliation[MPIP]
{Max Planck Institute for Polymer research, Ackermannweg 10, 55128 Mainz, Germany}
\author{Luciano Colombo}
\affiliation[UniCA]
{Dipartimento di Fisica, Universit\`a di Cagliari, Cittadella Universitaria, I-09042 Monserrato (Ca), Italy}
\author{Davide Donadio}
\email{ddonadio@ucdavis.edu}
\affiliation[UCDavis]
{Department of Chemistry, University of California Davis, One Shields Avenue, Davis, California 95616, United States}

\title{%
Energy Relaxation and Thermal Diffusion in IR pump\-/probe Spectroscopy of hydrogen-bonded Liquids}


\begin{document}

\begin{abstract}

Infrared pump\-/probe spectroscopy provides detailed information on the dynamics of hydrogen\-/bonded liquids.
Due to dissipation of the absorbed pump pulse energy, also thermal equilibration dynamics contributes to the observed signal. 
Disentangling this contribution from the molecular response remains an open challenge. Performing non\-/equilibrium molecular dynamics simulations of liquid deuterated methanol, we show that faster molecular vibrational relaxation and slower heat diffusion are decoupled and occur on different length scale. 
Transient structures of the hydrogen bonding network influence thermal relaxation by affecting thermal diffusivity over the length\-/scale of several nanometers. 
\end{abstract}

The transport of excess energy is an essential feature that dictates the performance of a liquid solvent as reaction medium, since the dissipation of excess energy in exothermic reactions can drive chemical conversion\cite{Marcus}. 
Hydrogen\-/bonded liquids are particularly efficient at transferring excess energy through strong coupling. Such energy transfer can be traced using infrared (IR) pump\-/probe experiments: excitation of specific, IR active oscillators, is used to label this very oscillator and monitor its fast dynamics as a function of time, frequency, or orientation\cite{Hamm}. 
Strong anharmonic coupling in hydrogen\-/bonded liquids leads to rapid vibrational relaxation, resulting eventually in a thermal distribution of the excess energy over lower frequency states\cite{Stingel}. However, the relevant length and timescale, as well as the relevant vibrational modes involved in this equilibration have so far remained elusive.

In deuterated liquids the relaxation dynamics is probed by the spectral changes of the OD stretching mode\cite{Mazur,Deng}. Pump\-/probe IR experiments on deuterated alcohols and water shows that, after excitation, the hydroxyl stretching mode relaxes on a fast sub-picosecond timescale, and eventually the excess energy equilibrates over longer timescale of the order of several picoseconds\cite{Mazur,Lock,Lesnicki:2018cf}. At the end of this process the vibrational spectrum corresponds to that of a system at slightly higher temperature. The relaxation to this hot ground state is commonly observed in third-order IR experiments\cite{Mazur, Dettori4, Yagasaki1,Cowan,Lock2,Nienhuys1,Huse,Ashihara1,Ashihara2,Lindner1,Lindner2,Bakker2,Fecko2,Kropman,Chieffo,Steinel,Schafer,Liu,Hunger}
The decay of the orientational memory takes up to 30 ps and depends on the average distance among the excited OD groups. This energy transfer process has been described using a continuum thermal relaxation model\cite{Liu, Hunger, Seifert}, in  the locally heated hydroxyl is approximated by a sphere of radius $R$ with temperature $T+\Delta T$, while the surrounding bath has a temperature $T$. 
This model describes correctly the trends of energy relaxation, but it accounts quantitatively for the observed relaxation of the orientational memory only if one assumes a thermal diffusivity ($\bar\kappa$) much lower than that of the liquid\cite{Liu,Mazur}.  For example, in the case of methanol the model yields a thermal diffusivity eight times lower than the macroscopic $\bar{\kappa}^{exp.}=10.1$ \AA{}$^2$/ps\cite{Williams}. Such discrepancy may rise from neglecting the differences in the vibrational modes of the hydroxyl group and the molecular details of the physical process at the characteristic size scale of the distance between hydrogen\-/bonded molecules, i.e. few Angstrom.

In this Letter we show that the energy relaxation following the excitation of the OD stretching mode of deuterated methanol (CH$_3$OD) occurs in two stages, which entail distinct relaxation times. 
By utilizing spectrally-resolved non\-/equilibrium molecular dynamics simulations, we characterize a rapid molecular energy relaxation with a timescale of $\sim10$ ps, which is intertwined with thermal diffusion, with a slower relaxation time that depends on the size of the system.
In sufficiently  large simulation cells the two processes are decoupled and the thermal relaxation time is dictated by the bulk thermal diffusivity.   

Methanol was modeled using the fixed\-/charge COMPASS force\-/field\cite{Sun}. Charges were obtained by fitting the electrostatic potential of an all electron Hartree\-/Fock calculation with the restrained electrostatic potential method\cite{Bayly,Cornell}. Long\-/range Coulomb interactions were computed using the particle\-/particle particle\-/mesh solver\cite{Hockney}. All the simulations have been performed using the LAMMPS package\cite{Plimpton}, in which the equations of motion are integrated with a timestep of 0.5 fs. The models have an average equilibrium density of $\rho=0.796$ g/cm$^{3}$, in excellent agreement with the experimental value at room temperature and normal pressure, $\rho_{exp}=0.81$ g/cm$^{3}$\cite{Oneil}.

To simulate the excitation in IR pump\-/probe experiments we applied a recently developed approach, based on the Generalized Langevin equation (GLE)\cite{Ceriotti1,Ceriotti2,Ceriotti3,Ceriotti4,Ceriotti5,Dettori4}.
GLE is non\-/Markovian extension of the Langevin equation, in which history\-/dependent features\cite{Zwanzig} are introduced by replacing the friction term with a convolution with a suitable memory kernel $K(t)$\cite{Nyquist,Callen}. Such memory kernel can be chosen to heat up the vibrational modes within a range $\Delta\omega$ of a chosen frequency ($\omega_{\rm peak}$), while the remaining modes are kept at a baseline temperature ($T_{\rm base}$). This implementation of  {\sl hot\-/spot} GLE consists of the combination of a standard white\-/noise Langevin thermostat with friction $\gamma_{\rm base}$ and target temperature $T_{\rm base}$ and a $\delta$\-/thermostat\cite{Ceriotti4}, at $T_{\rm peak}$ that is coupled with a friction parameter $\gamma_{\rm peak}$.\cite{Dettori4}
The resulting memory kernel in the frequency domain reads as:
\begin{equation}
\label{eq:deltakernel}
	  K(\omega) =  2\gamma_{\rm base} + \dfrac{\gamma_{\rm peak}}{\pi}
	  \dfrac{ \omega_{\rm peak}\Delta\omega\omega^2}{(\omega^2-\omega_{\rm peak}^2)^2+\Delta\omega^2\omega^2}.
\end{equation}
This approach has proven to be a reliable tool to model the relaxation dynamics of HBLs, such as methanol and water, upon pump\-/probe IR spectroscopy, providing for methanol relaxation times in excellent agreement with experiments\cite{Dettori4}. However, these preliminary simulations were carried out by exciting the whole system homogeneously, and they cannot provide insight into thermal diffusion. 

In order to shed light into both vibrational relaxation and thermal diffusion we use {\sl hot\-/spot} GLE within an approach to equilibrium MD (AEMD) setup\cite{Melis1}. 
AEMD is customarily used to compute the thermal diffusivity of a system by probing the transient regime occurring between an initial non\-/equilibrium configuration and the final equilibrium state, reached by integrating the equations of motion at constant energy and constant volume. The initial state is prepared such that half of the simulation cell is set at high temperature ($T_H$) and the other half at low temperature ($T_L$), as shown in Fig. S2. The temperature difference is usually maintained by two white\-/noise (e.g. Langevin) thermostats.   

When the thermostats are switched off, the system evolves toward equilibrium according to the one\-/dimensional heat equation with periodic boundary conditions: 
\begin{equation}
\label{eq:heatequation}
\dfrac{\partial T}{\partial t}=\bar{\kappa}\dfrac{\partial^2 T}{\partial z^2}.
\end{equation}
According to this equation the average temperature difference between the two regions of the system $\Delta T(t)=\langle T_H\rangle - \langle T_H\rangle$ evolves as 
\begin{equation}
\label{eq:deltat}
\Delta T(t)=\sum_{n=1}^\infty C_n e^{-\alpha^2_n\bar{\kappa}t}
\end{equation}
where $\alpha_n=2\pi n /L_z$, $\bar{\kappa}$ is thermal diffusivity, that serves to fit the computational data, and the coefficients $C_n=8(T_1-T_2)\dfrac{[cos(\alpha_nL_z/2)-1]^2}{\alpha_n^2 L_z^2}$ include information on the geometry and on the initial conditions.
As all the terms except $\bar\kappa$ in Eq. \eqref{eq:deltat} are known, fitting $\Delta T(t)$ to an exponentially decaying function provides $\bar\kappa$. For the sake of comparison and to validate our model, we calculate the thermal diffusivity of our deuterated methanol model at room temperature both by equilibrium MD and by AEMD. The thermal diffusivity obtained by equilibrium MD is in excellent agreement with the one determined experimentally for methanol $\bar{\kappa}^{exp.}=10.1$ \AA{}$^2$/ps\cite{Williams}. In AEMD simulations $\bar\kappa$ is calculated as  $\bar{\kappa}=\kappa/\rho c_v$ with $c_v=85.8$ J/mol K\cite{Katayama} and exhibits a strong size dependence, but it can be extrapolated to a value compatible with experiments (see Figure S2).

To study the energy relaxation processes after a colored laser excitation we prepare the ``hot'' half of the CH$_3$OD simulation box using the ``hot\-/spot'' GLE thermostat, parameterized to excite the OD stretching bond ($\omega_{\rm peak}=2477$ cm$^{-1}$), while the ``cold'' half of the system is kept at $T_0 \equiv T_{\rm base}=300$ with a standard Langevin bath. 

The excitation parameters of the GLE thermostat were chosen consistently with the previous GLE application\cite{Dettori4}: the colored noise contribution of the thermostat was characterized by $T_{\rm max}=400$~K, $1/\gamma_{\rm peak}=0.5$~ps and $\Delta \omega=1$~${\rm cm}^{-1}$, while the white noise Langevin part was characterized by $T_{\rm base}=300$~K and $1/\gamma_{\rm base}=0.5$~ps. 

Analogously to standard AEMD, after the thermostats are switched off,  we probe the transient regime by monitoring the evolution of the temperature. 
However, since spectroscopy experiments probe spectral changes directly related to the deuterated hydroxyl groups, we  monitor the difference between the deuterium temperature averaged over the excited half of the sample and the deuterium temperature averaged over the unperturbed half: 
\begin{equation}
\label{eq:deltaTdeuterium}
\Delta T^D(t)=\langle T^D_H(t) \rangle - \langle T^D_C(t) \rangle
\end{equation}
This choice allows us to reduce the thermal noise due to the other unperturbed molecular degrees of freedom:
\begin{figure}[t]
\begin{center}
\includegraphics[scale=0.25]{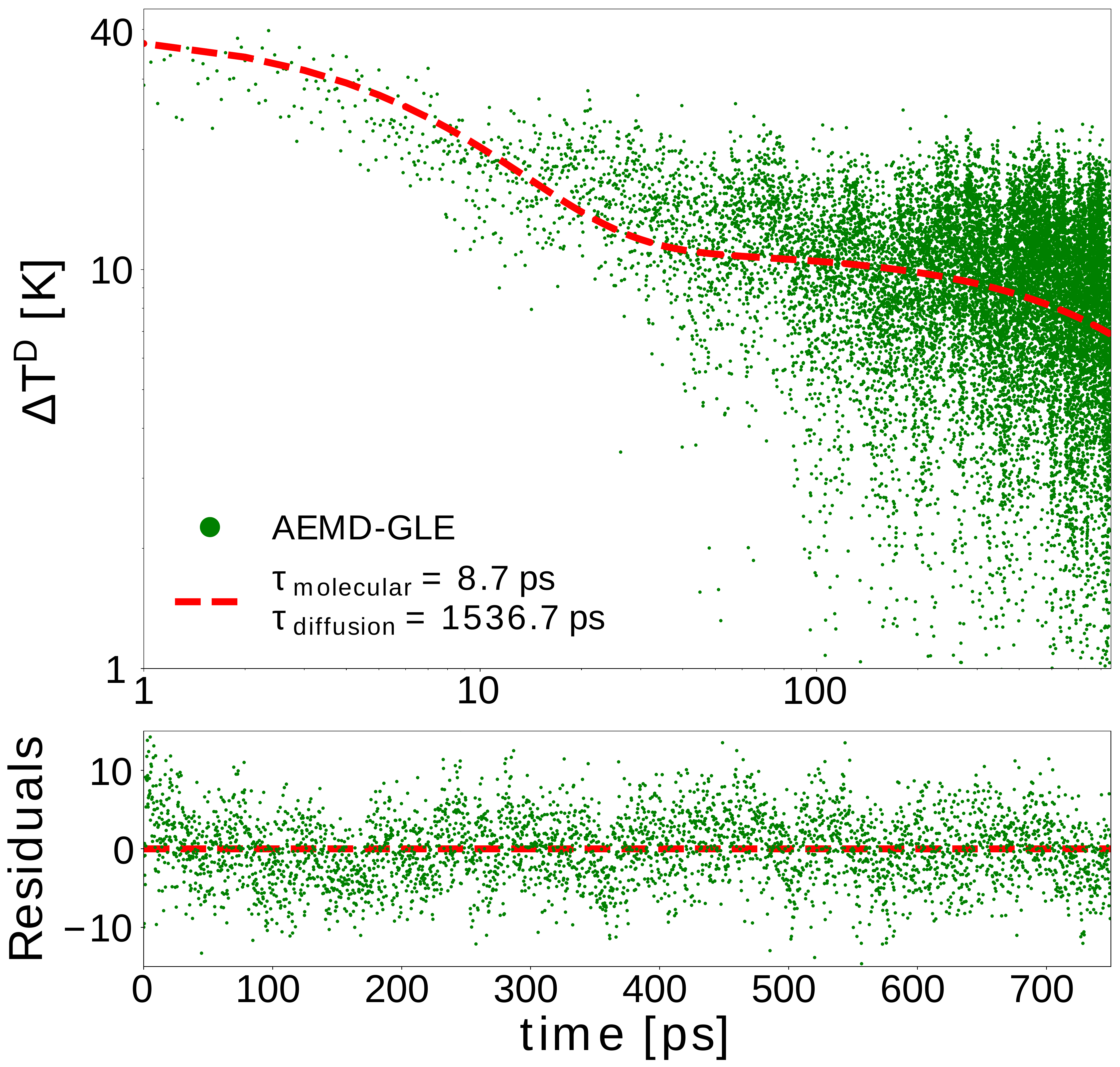}
\caption{Temperature difference in a log-log plot as a function of time for a deuterated methanol model, containing $38880$ atoms, with $L_z=74.33$ nm and a section of $2.44\times2.44$ nm$^2$. The red dashed line is a fit performed with the bi\-/exponential model of Eq. \eqref{eq:deltaT_decay}, considering $\tau_{\text{molecular}}=8.7$~ps, which produces $\tau_{\text{diffusion}}=1536.7$~ps and $\beta=0.73$. The lower panel shows the plot of the residuals, which are scattered around zero and confirm the appropriateness of the regression model here adopted.}
\label{fig:deltaTD_decay}
\end{center}
\end{figure}

Figure~\ref{fig:deltaTD_decay} shows the time dependence of the temperature difference for a model of CH$_3$OD made of $6480$ molecules with $L_z=74.33$ nm. The trend is characterized by an initial fast relaxation followed by a slower decay. In contrast to what happens in a standard AEMD simulation\cite{Melis1}, in this transient regime  $\Delta T^D(t)$ cannot be fitted by a single exponential decay, but it requires a double exponential form with two different relaxation times, suggesting the presence of two distinct energy relaxation mechanisms: 
\begin{equation}
\label{eq:deltaT_decay}
\Delta T^D(t)=\Delta T^D_0\left[\beta e^{-t/\tau_{\text{molecular}}}+(1-\beta) e^{-t/\tau_{\text{diffusion}}}\right].
\end{equation}
Here $\Delta T_0^D$ is the temperature difference at $t=0$ computed for the deuterium atoms between the two regions, and $\tau_{\text{molecular}}$ and $\tau_{\text{diffusion}}$ are the two relaxation times. 
The bi\-/exponential trend is observed in all our AEMD-GLE simulation, regardless of the system size. 

pump\-/probe experiments suggest that after the excitation a sub\-/picosecond quantum relaxation occurs, which cannot be probed by classical simulations\cite{Mazur}. After this fast process the system is in a non\-/equilibrium state that in our model corresponds to that prepared with the {\sl hot\-/spot} GLE approach. This state decays over a characteristic molecular timescale of several picoseconds, corresponding to $\tau_{molecular}$, after which the formerly excited molecules can be considered in a vibrational ground state with a higher molecular temperature $T^\star>T_{\text{eq}}$. 
Such excess heat is then transferred to the surroundings so to achieve thermal equilibrium. This process is controlled by thermal diffusion, which determines the longer relaxation time $\tau_{\text{diffusion}}$. The biexponential decay in Eq.~\ref{eq:deltaT_decay}  supports this interpretation.
In addition, the $\beta$ parameter in Eq.~\ref{eq:deltaT_decay} provides an estimate of the contribution to the overall relaxation process in terms of the energy carried by the two decay processes: the kinetic temperature of the atomic species accounts for both the energy that excited OD stretching bonds transfer to the low frequency OD modes and the excess heat dissipated via thermal diffusion.

The relaxation time computed in these AEMD\-/GLE simulations is $\tau_{molecular}=8.7\pm0.5$ ps, in agreement with our former simulations in which the entire model was homogeneously excited with {\sl hot\-/spot} GLE\cite{Dettori4}. This process corresponds to a rapid localized rearrangement of the ``colored'' energy from OD stretching to the bending ($\omega\sim 900$ cm$^{-1}$) and librational ($\omega\sim 500$ cm$^{-1}$) modes of the deuterated hydroxyl group. 

The longer decay time ($\tau_{\text{diffusion}}$) corresponds to 1.5~ns and the weight $\beta$ is 0.73. We hypothesize that $\tau_{\text{diffusion}}$  is related to thermal diffusion. Within this assumption we can make use of the one\-/dimensional solution of the heat equation with periodic boundary conditions (Eqs.\eqref{eq:heatequation}--\eqref{eq:deltat}) to extract a thermal diffusivity  $\bar\kappa=\left(2\pi/L\right)^2/\tau_\mathrm{diffusion}$.\footnote{Here we have assumed that the series in \eqref{eq:deltat} is dominated by the fundamental component ($n=1$).}
For the system showcased in Figure~\ref{fig:deltaTD_decay} $\bar\kappa=$9.1 \AA{}$^2$/ps is in very good agreement with the experimental diffusivity and with our Green\-/Kubo calculation. 

\begin{figure}[t]
\begin{center}
\includegraphics[scale=0.21]{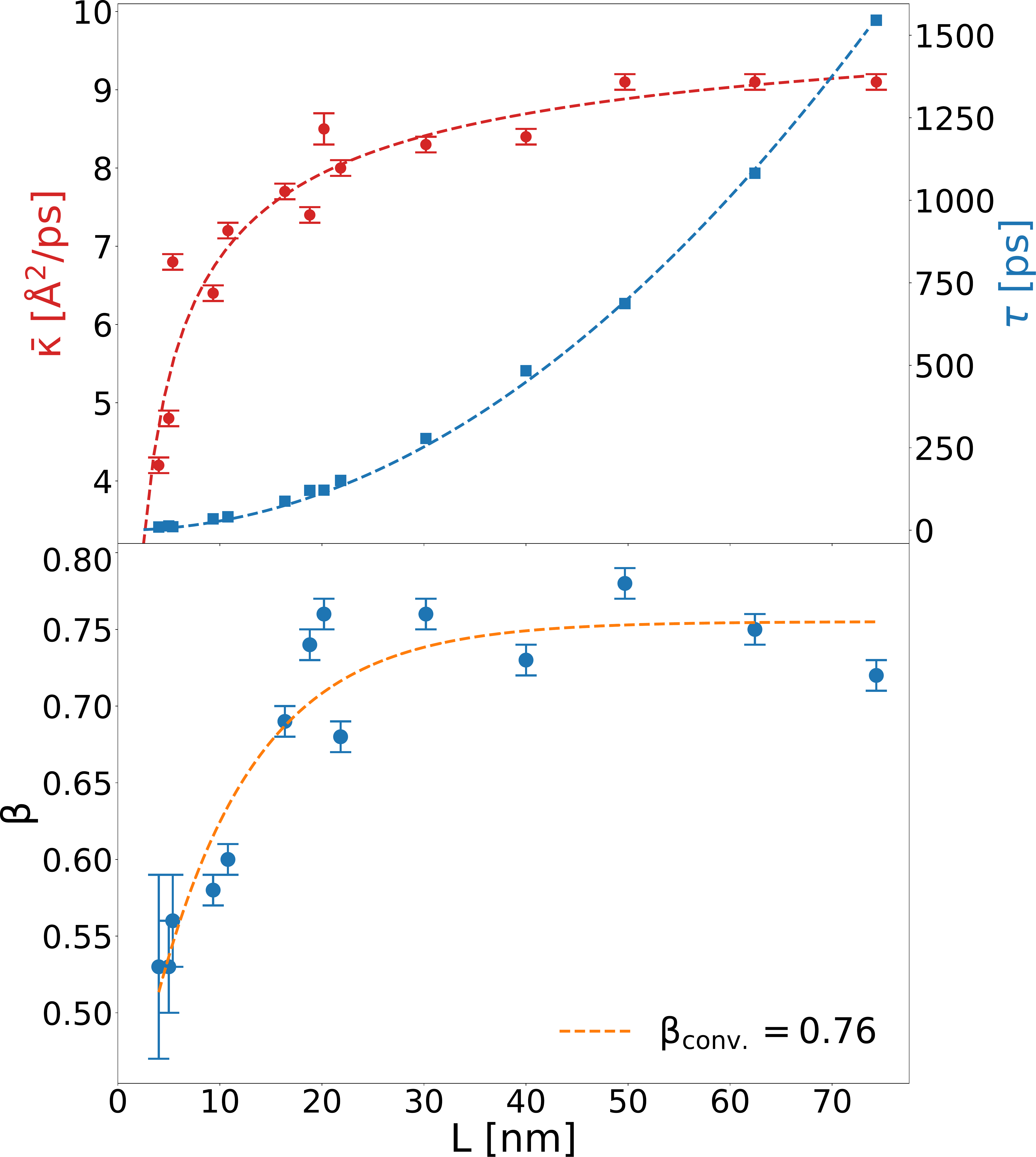}
\caption{Top panel. Heat diffusion times $\tau_{\text{diffusion}}$ (blue squares, right scale) and the corresponding diffusivities (red circles, left scale) computed with the respective system length $L$ as a function of system size. Bottom panel. $\beta$ parameters obtained by means of Eq. \eqref{eq:deltaT_decay} as a function of system length.  An exponential fit (dashed orange curve) was used to estimate $\beta=0.76$ and a characteristic decay length of $\sim11$ nm.}
\label{fig:GLEdiffusivity}
\end{center}
\end{figure}

To provide further evidence that the slow relaxation time is related to thermal diffusion, we applied the AEMD\-/GLE protocol to several simulation cells, with length $L_z$ between $4.0$ and $74.3$~nm. Results are obtained by averaging over 160 independent simulations. $\tau_\mathrm{diffusion}$ and the corresponding values of thermal diffusivity as a function of $L_z$ are shown in Figure~\ref{fig:GLEdiffusivity}(top panel). These two quantities increase with the size of the system, as observed also in the plain AEMD case (see SI). $\bar\kappa$ saturates after $\sim40$ nm to a limit value similar to the bulk $\bar\kappa$ computed by equilibrium MD.
These trends of $\tau_{\text{diffusion}}(L_z)$ and $\bar\kappa (L_z)$ support the hypothesis that $\tau_{\text{diffusion}}$ is dictated by thermal diffusion.
The low values of $\bar\kappa (L_z)$ for $L_z<20$~nm suggest that below this characteristic length\-/scale molecular relaxation and thermal diffusion are entangled, whereas for larger systems the time scales are so different that the two processes are completely decoupled.

We analyze the relative weight of the two relaxation mechanisms by plotting the parameter $\beta$ in Eq. \ref{eq:deltaT_decay} as a function of $L_z$ (Fig.\ref{fig:GLEdiffusivity}, bottom panel). The size dependence of $\beta$ suggests that the two relaxation mechanisms contribute in different proportions at different length scales. An exponential relation was used to calculate the asymptotic value of $\beta_{\rm conv}=0.76$ at large scale.
The entanglement observed at short length scales implies that thermal diffusion occurs while the vibrational excitation has not yet fully relaxed. Heat transfer in these conditions is limited by the small distance among groups of molecules, in which OD\-/related vibrational modes are overpopulated following the initial excitation.

In order to gain molecular insight and a clearer picture of the energy relaxation process, we compare time\-/dependent spectra, obtained by performing short time Fourier Transform (STFT)\cite{Allen3} of the velocity\-/velocity autocorrelation function:
\begin{equation}
    I(t,\omega)=\int_{-\infty}^{\infty}\langle\mathbf{v}(t)\mathbf{v}(t+\tau)\rangle f(t+\tau) e^{-i\omega \tau}d\tau \ .
\end{equation}
 $f(t)=\sum_{i=0}^{L}\alpha_i cos\left(i \frac{2\pi}{N}t\right)$ is a window function, which is nonzero only for a short time interval, with $N$ total number of timesteps and $\alpha_i$ coefficients conveniently chosen to reduce spectral leakage and avoid spurious components in the frequency spectrum. We chose the Blackman\-/Nuttall window function with $L=7$  in order to preserve the correct amplitude of the vibrational spectra. 
\begin{figure}[t]
\begin{center}
\includegraphics[scale=0.2]{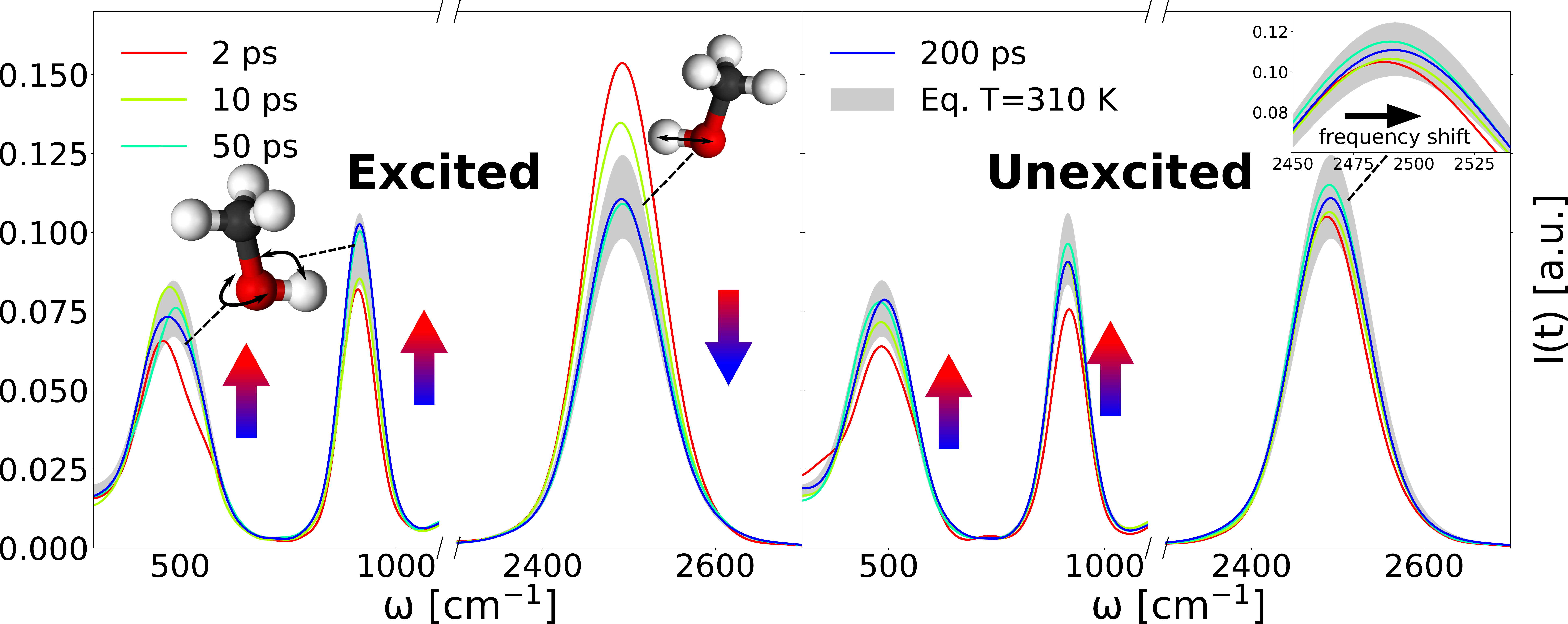}
\caption{Time\-/dependent vDOS obtained for the excited and the unexcited region of the $74.3$ nm\-/long simulation cell. The grey shaded area represents a standard deviation interval on the equilibrium vDOS averaged over 10 independent replicas and calculated with respect to the final configuration at T=310 K. For the sake of clarity, we excluded the frequencies between 1100 and 2300 cm$^{-1}$ since these modes do not contribute to energy relaxation\cite{Dettori4}. }
\label{fig:tvdos}
\end{center}
\end{figure}
Figure \ref{fig:tvdos} displays the comparison of the time\-/dependent vibrational density of states (vDOS) obtained for the excited (left) and  unperturbed (right) regions of the sample with $L_z=74.33$ nm. Here the grey shaded area represents the uncertainty of the equilibrium vDOS averaged over 10 independent replicas of a system at T=310 K, i.e. the temperature attained after equilibration of the OD excitation. 
Left and right vDOS are obtained by excluding atoms at the boundaries between the excited and unperturbed regions, so to exclude mutual influence and spurious effects. 
In the interpretation of these time\-/resolved spectra we need to consider that MD is classical, so we do not expect quantum bleaching of the peaks, and changes in the intensity of the peaks are directly proportional to changes in the population of the corresponding vibrational modes.
The left\-/hand panel shows that 2 ps after excitation the OD stretching peak is overpopulated (red curve). Its intensity decays during the first 10 ps (light green curve) and it oscillates within the equilibrium range at later times (light blue and blue curves). 
The intensity of the bending and librational modes involving OD group increases over the same $\sim 10$ ps time scale, as they provide the faster channels of relaxation of the energy absorbed by OD stretching. This behavior confirms previous results on the energy transfer between  excited OD stretching and unperturbed modes of the OD group\cite{Dettori4}.

The the time\-/dependent vDOS in the right\-/hand panel accounts for the trends observed in Fig. \ref{fig:GLEdiffusivity}. 
The intensity of the OD stretching peak at $\omega=2477$ cm$^{-1}$ does not change significantly either in the first $\sim$10 ps or at longer times. However, the overall temperature increase shifts the center of the OD stretching band toward higher frequencies, as a consequence of the average weakening of the hydrogen bonds and a stiffening of the OD bond\cite{Dettori4,Mazur,Mazur2}. 
The intensity of the bending and librational peaks exhibits a sharp increase in the first 10~ps in a similar fashion as in the excited region, implying that energy redistributes non\-/thermally across the whole sample. Since the energy transfer from OD stretching to these modes is highly influenced by hydrogen bonds, we assume that this fast non\-/thermal energy redistribution is mediated by the hydrogen bonded network through structures that extend across the excited region and the unexcited one.
Eventually the energy redistributes to the other degrees of freedom through thermal diffusion.

The spatially resolved analysis of transient vDOS highlights the effect of the hydrogen bond network on the fast energy relaxation process.  
To further understand the interplay between the two relaxation time scales, especially on small length scales, we characterize the mid\-/range structure of the hydrogen bonding network of methanol. To this end we performed Grazing\-/Incidence Wide\-/Angle X\-/ray Scattering (GIWAXS) simulations, following the procedure described in\cite{Coleman} and implemented in LAMMPS, assuming an incident X\-/rays have wavelength $\lambda = 1.54$ \AA. 
\begin{figure}[t]
\begin{center}
\includegraphics[scale=0.25]{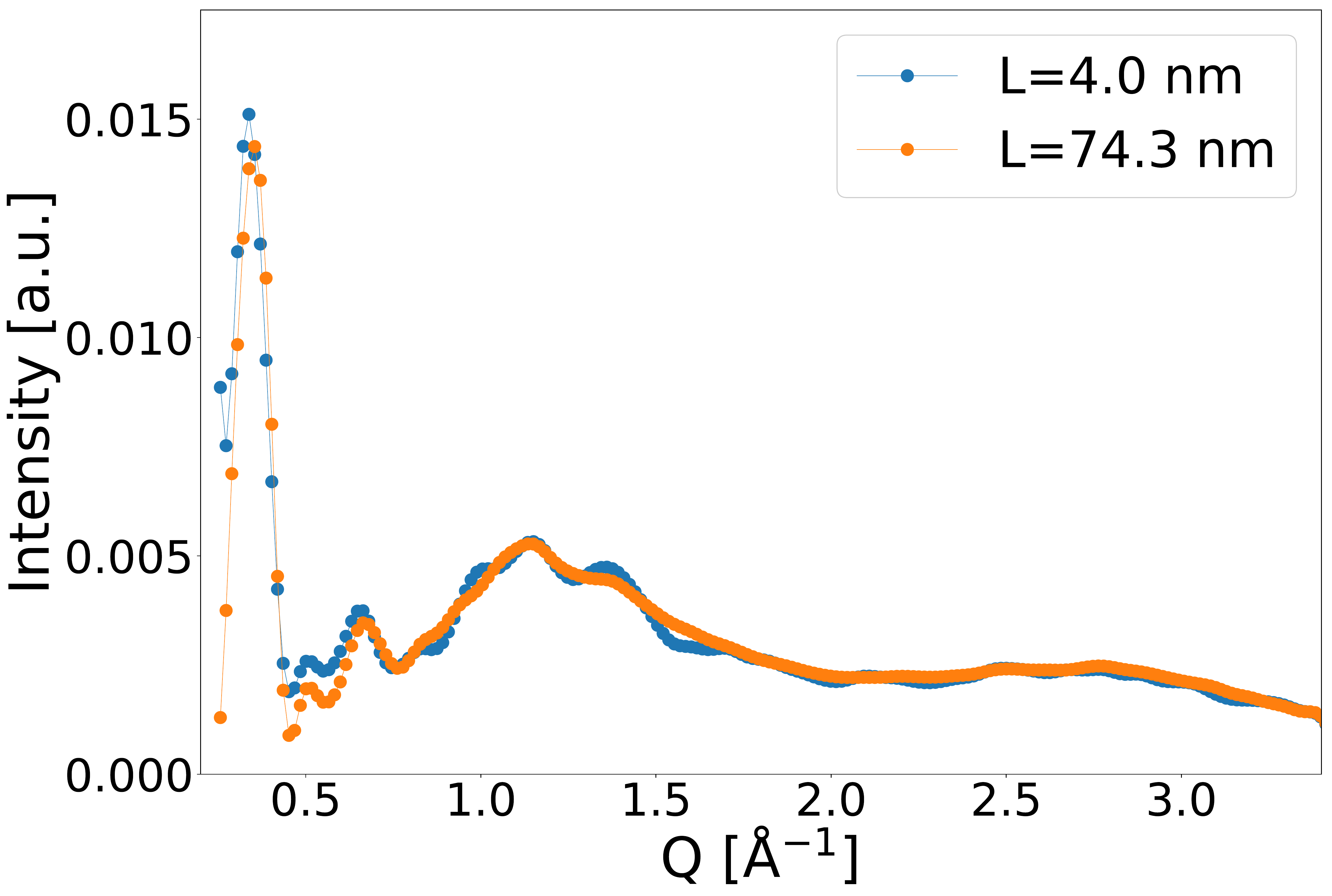}
\caption{Simulated x\-/ray diffraction pattern for the smallest (blue dots) and the largest (orange dots) simulation cell.}
\label{fig:xrd}
\end{center}
\end{figure}
The spectra in Fig. \ref{fig:xrd} refer to the smallest and the largest systems considered. The broad bands at $2.75$ and $1.25$ \AA{}$^{-1}$ account for the first and second nearest neighbor shells of hydrogen bonded molecules with characteristic lengths of $r=2\pi/Q=2.3$ \AA, and $r=2\pi/Q=5.2$ \AA, respectively. These peaks are consistent with published {\sl ab initio} oxygen\-/oxygen radial distribution functions.\cite{Pagliai,Handgraaf:2004es} .
Two further peaks appear in both simulation cells: a weak one, at $\sim 0.65$ \AA{}$^{-1}$ and a very intense one centered at $0.35$ \AA$^{-1}$. The latter indicates the presence of extended intermolecular structures $\sim 1.8$ nm\-/long.
Similar results were obtained by neutron scattering experiments and confirmed by atomistic simulations, where structures made of 14\-/15 molecules were observed with persistent length up to $14$ \AA, suggesting that methanol, and alcohols in general, entail medium range order\cite{Adya,Bertrand,Vrhovsek}. The presence of such structures correlates with the observed trends in the relative weight between the fast and the thermal energy relaxation processes, and for the apparently reduced thermal diffusivity on short length scales (Fig. \ref{fig:GLEdiffusivity}).  
The presence of these structures accounts for the occurrence of the fast molecular energy relaxation process in the initially unperturbed area of the system.  

In conclusion, our work provides a clear molecular picture of energy relaxation in IR pump\-/probe spectroscopy of methanol. We identify two distinct mechanisms, which can be either coupled or decoupled, depending on the characteristic length scale of the system. Specifically, we resolve a fast non\-/thermal relaxation that redistributes the excess population of OD stretching to OD bending and librations, and a slower relaxation that can be suitably described as thermal diffusion. 
When the two mechanisms are decoupled the relaxation time related to thermal diffusion depends on the length\-/scale of the system according to the heat equation. At short length\-/scales the two mechanisms are intertwined, and the systems exhibit reduced thermal diffusivity, as it is normally observed in quasi\-/ballistic nanoscale heat transport\cite{HoogeboomPot:2015gd}.
Both mechanisms are mediated by the persistent medium\-/range structures of the hydrogen bonding network of the system. 
Therefore, experiments that probe vibrational energy decay in isotopic mixtures, in which intermolecular distances among excited molecules can be determined by the ratio between deuterated and non\-/deuterated species, provides a fine fingerprint of the structure and dynamics of alcohols and hydrogen\-/bonded liquids in general. 
\begin{acknowledgement}
We are grateful to Mischa Bonn and Claudio Melis for useful discussions. 
\end{acknowledgement}

\begin{suppinfo}

Simulation cell set up for AEMD-GLE calculations and calculated thermal diffusivities for the whole set of simulation cells, both obtained with equilibrium and non-equilibrium molecular dynamics methods.

\end{suppinfo}

\providecommand{\latin}[1]{#1}
\makeatletter
\providecommand{\doi}
  {\begingroup\let\do\@makeother\dospecials
  \catcode`\{=1 \catcode`\}=2 \doi@aux}
\providecommand{\doi@aux}[1]{\endgroup\texttt{#1}}
\makeatother
\providecommand*\mcitethebibliography{\thebibliography}
\csname @ifundefined\endcsname{endmcitethebibliography}
  {\let\endmcitethebibliography\endthebibliography}{}

\end{document}